\documentstyle[12pt]{article}

\newcommand{\doublespace}{
\renewcommand{\baselinestretch}{1.6}\large\normalsize}
 
\begin{document}

\hyphenation{plaq-uette}

\begin{titlepage}

\begin{tabbing}
\`hep-lat/9605045\\
\`OUTP-96-26P \\
\` May, 1996 \\
\end{tabbing}
 
\vspace*{1.0in}
 
\begin{center}
{\bf Monopoles At Finite Volume and Temperature in
SU(2) Lattice Gauge Theory\\}
\vspace*{.3in}
John D. Stack \\
\vspace*{.1in}
{\it Department of Physics, \\
University of Oxford,  \\
1 Keble Rd,\\
Oxford OX1 3NP, U.K.\\
and\\
 Department of Physics,$^{*}$ \\
University of Illinois at Urbana-Champaign, \\
1110 W. Green Street, \\
Urbana, IL 61801, U.S.A. \\}
\vspace*{.2in}
Roy J. Wensley\\
\vspace*{.1in}
{\it Department of Mathematical Sciences,\\
Saint Mary's College,\\
Moraga, CA 94575, U.S.A\\}
\vspace*{.2in}
Steven D. Neiman\\
\vspace*{.1in}
{\it  Department of Physics, \\
 University of Illinois at Urbana-Champaign, \\
1110 W. Green Street, \\
Urbana, IL 61801, U.S.A. \\}
\end{center}

$^{*}$permanent address

\end{titlepage}
\vfill\eject

\doublespace
\pagestyle{empty}

\begin{center}
{\bf Abstract}
\end{center}

We resolve a discrepancy between the  $SU(2)$ spacial string
tension at finite temperature, and the value obtained by monopoles
in the maximum Abelian gauge.  Previous work had incorrectly omitted
a term due to Dirac sheets.  When this term is included, the monopole
and full $SU(2)$ determinations of the spacial string tension agree
to within the statistical errors of the monopole calculation.
\vspace*{.5in}
 
\newpage
\pagestyle{plain}

	The string tension in pure $SU(2)$ lattice gauge theory has
	recently been calculated from monopoles, giving results that agree with the
	full $SU(2)$ string tension, to within the statistical errors of the
	monopole calculations~\cite{jssnrw}.  Given
	this success, it is disappointing that at finite temperature, monopoles apparently fail
	to explain the fundamental representation spacial string tension,
	$\sigma_s$. In the
	region of temperatures above the deconfining transition, the
	values for $\sigma_s$ obained from monopoles are too small, by an
	amount well outside of statistical errors, and the discrepancy
	increases with increasing temperature~\cite{js_lat94,suzuki}.

	In this paper we resolve this discrepancy.
	In the part of the calculation in which Wilson loops are
	calculated from monopoles, a term involving Dirac sheet
	variables had been dropped. When this term
	is included as it should be, the monopole and full $SU(2)$ determinations of the
	fundamental spacial string tensions agree to within 
	statistical errors.
	The present calculations
	were carried out on lattices of size $16^{3}\times N_{t}$ at $\beta=2.5115$,
	for $N_{t}=4,6,8,12$.  At this value of $\beta$, the deconfining transition
	corresponds to $N_{t}=8$~\cite{fingberg}, so $N_{t}=4,6$ are in the high-temperature phase.
	We compare to the full $SU(2)$ results for $\sigma_s$  obtained by
	Bali {\it et al}~\cite{bali}
	at this same $\beta$ value on  $32^3\times N_{t}$ lattices. Before presenting our
	results, we briefly review the way the monopole calculations are done, and discuss the
	correction due to Dirac sheets.

	A summary of the steps~\cite{jssnrw} involved in the monopole calculations is
	as follows: (1) $SU(2)$ configurations are projected into the
	maximum Abelian gauge.  (2) The resulting $SU(2)$ links are
	factored into Abelian and charged parts. (3) The Abelian link
	angles are used to locate the magnetic current of the monopoles.
	(4) Monopole Wilson loops are calculated.
	(5) The string tension is extracted from fits to the
	monopole Wilson loops.  Steps (3)-(5) proceed exactly as they would
	in $U(1)$ lattice gauge theory.

	For $SU(2)$ lattice gauge theory, these steps involve strong assumptions.  
	Briefly, these are
	that in the maximum Abelian gauge (MAG), long range physics is
	Abelian and controlled by monopoles.  
	Each link of a configuration in
	the MAG is approximated by its Abelian part,
	$\exp(i\phi^{3}_{\mu}\tau_{3})$, where
	$\tau_{3}=\sigma_{3}/2$ 
	is the isospin generator, and $-2\pi \leq \phi^{3}_{\mu}
	\leq 2\pi$.  The key assumption for Wilson loops is
	$$
	<W_{SU(2)}>\: \sim \: <W_{\phi}>,
	$$
	where the $\sim$ sign means equivalent at long range. The
	Abelian Wilson loop $ W_{\phi}$ is given by
	\begin{equation}
	W_{\phi} = \exp(i\sum_{x,\mu}\phi_{\mu}(x)J_{\mu}(x)).
	\label{eqn_Wphi}
	\end{equation}
	In Eq.(\ref{eqn_Wphi}), $\phi_{\mu}$ is a rescaled link angle,
	$\phi_{\mu}=\phi^{3}_{\mu}/2$,
	while $J_{\mu}$ is an integer-valued current describing the path
	traversed by the heavy quark. Motivated by results in $U(1)$ lattice
	gauge theory~\cite{jsrw}, 
	the Abelian Wilson loop $W_{\phi}$ is further assumed to factor into a
	short range part involving the exchange of neutral gluons, times
	a term $W_{mon}$  arising from monopoles.  Thus for the confining part
	of the heavy quark potential 
	\begin{equation}
	<W_{SU(2)}> \: \sim \: <W_{\phi}> \: \sim \: <W_{mon}>
	\label{eqn_assume}
	\end{equation}
	is supposed to hold.  In particular,
	$W_{SU(2)}$, $W_{\phi}$, and
	$W_{mon}$ should all produce the full $SU(2)$ string tension.  
	Here we are concerned with the correct
	calculation of $W_{mon}$, and therefore with the second
	equivalence of Eq.(\ref{eqn_assume}), $<W_{\phi}>\  \sim \ <W_{mon}>$.
	The assumption of Abelian dominance and the use of the MAG are mainly concerned
	with the first equivalence in Eq.(\ref{eqn_assume}). 

	 To
	calculate $W_{mon}$,  monopole variables must be
	located in a $\phi_{\mu}$ configuration~\cite{degrand}.  Abelian plaquettes $\phi_{\mu\nu}$
	are formed from the link angles $\phi_{\mu}$, and expressed as
	$\phi_{\mu\nu}=\phi_{\mu\nu}^{\prime} + 2\pi m_{\mu\nu}$, where
	$\phi_{\mu\nu}^{\prime} \in [-\pi,\pi]$, and $m_{\mu\nu}$ is an integer.
	The surface formed by the dual variables $m^{*}_{\mu\nu}$ describes the Dirac sheets
	which are    
	present. \footnote{ Our conventions are that
	direct lattice link variables ($j_{\mu},A^{e}_{\mu}$), originate from the labelling site,
	while dual lattice link variables ($m_{\mu},A^{m}_{\mu}$) terminate at the the labelling site.  Direct
	lattice plaquette variables ($m_{\mu\nu},F^{e}_{\mu\nu},F^{m*}_{\mu\nu}$) 
	have the site at the lower left
	corner of the plaquette, while dual lattice plaquette variables 
	($m^{*}_{\mu\nu},F^{e*}_{\mu\nu},F^{m}_{\mu\nu}$)
	have the site  at the upper right corner. 
	In going from the direct to the dual lattice, the discrete difference
	$\partial^{\pm}_{\mu}$ is replaced by $\partial^{\mp}_{\mu}$, and vice versa.}
	Two currents can
	be derived from $m_{\mu\nu}$, one the magnetic current of the
	monopoles, $m_{\mu}= \partial_{\nu}^{+}m_{\mu\nu}^{*}$,
	the other the electric Dirac sheet
	current, $j_{\mu} = \partial_{\nu}^{-}m_{\mu\nu}$.  
	Geometrically, $m_{\mu}$ flows on the edge of
	the open surface defined by $m_{\mu\nu}^{*}$. 
	To obtain $W_{mon}$, we start from an expression  which can be derived analytically
	in (Villain) $U(1)$ lattice gauge theory \cite{jsrw},
	\begin{equation}
	W_{mon} = \exp(2\pi i \sum_{x,\mu} J_{\mu}(x) A^{e}_{\mu}(x)).
	\label{eqn_Wmon}
	\end{equation}
	Here $A^{e}_{\mu}$  is an electric vector potential whose source is $j_{\mu}$.
	The subtlety with which the present paper is concerned arises when an
	attempt is made to express $W_{mon}$ in terms of the magnetic current
	$m_{\mu}$.  This cannot be done using vector potentials, since the
	magnetic current produces a magnetic vector potential $A^{m}_{\mu}$,
	whereas  the electric current of the quark can only couple to an electric
	vector potential, in this case $A^{e}_{\mu}$.
	Progress can be made by going to field strengths.  We introduce
	a surface $D_{\mu\nu}$  of plaquettes whose boundary is the heavy quark
	current $J_{\mu}$, so that $J_{\mu} = \partial_{\nu}^{-}D_{\mu\nu}$.  
	Then using the lattice form of Stoke's theorem, $W_{mon}$  can be written
	as the exponential of a  flux integral
	\begin{equation}
	W_{mon} = \exp(\frac{2\pi i}{2} \sum_{x,\mu,\nu}D_{\mu\nu}(x)F_{\mu\nu}^{e}(x)),
	\label{eqn_flux}
	\end{equation}
	where the field strength is given by $F^{e}_{\mu\nu} \equiv \partial_{\mu}^{+}A^{e}_{\nu}
	- \partial_{\nu}^{+}A^{e}_{\mu}$, and satisfies the electric Maxwell equation,
	$\partial_{\nu}^{-}F^{e}_{\mu\nu}
	= j_{\mu}$.
	Analogous to $A^{e}_{\mu}$,  there is a
	magnetic vector potential $A^{m}_{\mu}$,  whose source is $m_{\mu}$.  The field
	strength given by $F^{m}_{\mu\nu} = \partial^{-}_{\mu}A^{m}_{\nu}-
	\partial^{-}_{\nu}A^{m}_{\nu}$, satisfies the magnetic Maxwell equation,
	$\partial^{+}_{\nu}F^{m}_{\mu\nu} = m_{\mu}$.

	 In a finite volume, it is not valid to replace $F^{e}_{\mu\nu}$ 
	in the exponent of Eq.(\ref{eqn_flux}) by  the dual of $F^{m}_{\mu\nu}$,  defined as usual by
	$ F^{m*}_{\mu\nu}= \frac{1}{2}\epsilon_{\mu\nu\alpha\beta}F^{m}_{\alpha\beta}$. 
	For the case of periodic boundary condition in all directions, the 
	correct
	equation relating $F^{e}_{\mu\nu}$  and $F^{m*}_{\mu\nu}$ is 
	\begin{equation}
	F^{e}_{\mu\nu}(x) + F^{m*}_{\mu\nu}(x) = m_{\mu\nu}(x) - \bar{m}_{\mu\nu},
	\label{eqn_dual}
	\end{equation}
	where $\bar{m}_{\mu\nu}$ is the space-time average of $m_{\mu\nu}$,
	\begin{equation}
	\bar{m}_{\mu\nu} = \frac{1}{V} \sum_{x} m_{\mu\nu}(x).
	\end{equation}

	It is straightforward to derive Eq.(\ref{eqn_dual}).  Here we simply list various
	consistency checks.  Applying $\partial^{-}_{\nu}$
	to Eq.(\ref{eqn_dual}) gives an identity 
	since $\partial^{-}_{\nu}F^{m*}_{\mu\nu}=0$ , and 
	$\partial^{-}_{\nu}F^{e}_{\mu\nu} = \partial^{-}_{\nu}m_{\mu\nu} = j_{\mu}$.
	Likewise, the equation obtained by applying $\partial^{+}_{\nu}$ to
	the dual of Eq.(\ref{eqn_dual}) is identically satisfied.  The
	constant  $-\bar{m}_{\mu\nu}$ on the right side of Eq.(\ref{eqn_dual}) is needed to
	make the space-time average of the right hand side vanish.  The
	space-time average of the left hand side vanishes for periodic
	boundary conditions,  since $F^{e}_{\mu\nu}$ and
	$F^{m*}_{\mu\nu}$  are linear combinations of gradients in their respective vector potentials,
	so the sum over all $x$ of $F^{e}_{\mu\nu} + F^{m*}_{\mu\nu}$ vanishes.

	Using Eq.(\ref{eqn_dual}) to replace $F^{e}_{\mu\nu}$  in Eq.(\ref{eqn_flux}), we have
	\begin{equation}
	W_{mon} = \exp(-\frac{2\pi i}{2} \sum_{x,\mu,\nu} D_{\mu\nu}(x)(F^{m*}_{\mu\nu}(x)
	+\bar{m}_{\mu\nu})).
	\label{eqn_dual_flux}
	\end{equation}
	The integer $m_{\mu\nu}$ term on the right hand side of Eq.(\ref{eqn_dual}) 
	has no effect on $W_{mon}$. However, the presence of the non-integer term $\bar{m}_{\mu\nu}$ 
	in Eq.(\ref{eqn_dual_flux}) means that in addition to the magnetic current,	
	the six numbers $\bar{m}_{\mu\nu}$ must be specified to obtain $W_{mon}$.
	Since Eq.(\ref{eqn_dual_flux}) is 
	equivalent to Eq.(\ref{eqn_flux}), and therefore to Eq.(\ref{eqn_Wmon}), the necessary
	property that $W_{mon}$ is unity for a plane-filling loop is now guaranteed.  This
	is false if the $\bar{m}_{\mu\nu}$ term is omitted in the exponent of Eq.(\ref{eqn_dual_flux}).
	While these three forms for $W_{mon}$ are completely equivalent, there are reasons
	for preferring the
	representation of Eq.(\ref{eqn_dual_flux}).  
	If monopoles are the cause of non-perturbative phenomena, it is desirable
	to express physical quantities in terms of $m_{\mu}$ to the greatest
	extent possible.
	Further, the
	magnetic current is sparse, occupying only a small percentage of the
	links of the lattice,
	 and is essentially limited to the values
	$\pm 1$.  The electric Dirac sheet current, $j_{\mu}$, on the other hand,
	is dense, occupying a large fraction of the links of the
	lattice, and is not dominated by the values $\pm 1$.
	Finally, $j_{\mu}$  varies under
	local deformations of Dirac sheets.
	The deformation
	$m_{\mu\nu} \rightarrow m_{\mu\nu} + \partial^{+}_{\mu}n_{\nu}
	-\partial^{+}_{\nu}n_{\mu}$, $n_{\mu}$ an integer,
	leaves $m_{\mu}$  and $\bar{m}_{\mu\nu}$  unchanged, while $j_{\mu}$ does change,
	$j_{\mu} \rightarrow j_{\mu} + \partial^{+}_{\mu} (\partial^{-} \cdot n)
	- (\partial^{+} \cdot \partial^{-}) n_{\mu}$. 
	Although $W_{mon}$ is invariant
	under local Dirac sheet deformations, 
	there remains  a dependence 
	on the overall Dirac sheet
	topology.  For example, a 
	loop of magnetic current  could have a Dirac sheet 
	consisting of plaquettes with non-vanishing $m^{*}_{\mu\nu}$, which tile  the ``inner" area of
	the loop, or by virtue of the periodic boundary conditions,
	the ``outer" area.  The difference between the value of $W_{mon}$ for these two
	cases comes entirely from $\bar{m}_{\mu\nu}$.

\begin{table}
\begin{center}
\begin{tabular}{||c|c|c|c||} \hline
$N_{t}$ & $\sigma_{s}(m_{\mu})$ & $\sigma_{s}(m_{\mu},\bar{m}_{\mu\nu})$ & $\sigma_{s}(SU(2))$ \\ \hline
4  & 0.049(3) & 0.065(3) & 0.0643(6) \\ \hline
6  & 0.026(1) & 0.040(4) & 0.0381(4)\\ \hline
8  & 0.022(2) & 0.031(3) & 0.0325(7)\\ \hline
12 & 0.029(1) & 0.034(1) & -------- \\ \hline
\end{tabular}
\end{center}
\begin{center}
\bf{Table I}
\caption {Spacial string tensions from magnetic current, magnetic current + Dirac sheets,
and for full $SU(2)$}
\end{center}
\end{table}

	We now turn to our results. The methods we used for generating $SU(2)$ configurations
	and our gauge-fixing criterion are described in detail in~\cite{jssnrw}. 
	After equilibration, we saved configurations every 20 lattice updates, resulting in a total of 
	500 configurations of 
	magnetic current $m_{\mu}$, and the six Dirac sheet space-time averages, $\bar{m}_{\mu\nu}$ \cite{gribov}.
	As  we previously found at zero temperature, the magnetic current is sparse.  The average fraction of links
	carrying magnetic current can be conveniently written as $f \times 10^{-2}$.
	For our lattices of size $16^{3} \times N_{t}$ 
	the values of $f$ were $1.30(1),\: 1.07(1),\: 1.18(1),\: 1.25(1)$, for  $N_{t}=4,6,8,12$ respectively.
	The averages of $\bar{m}_{\mu\nu}$ are statistically zero, so as a measure of the effect of
	$\bar{m}_{\mu\nu}$ in Eq.(\ref{eqn_dual_flux}), we give $2\pi$ times the standard deviation
	of $\bar{m}_{\mu\nu}$, over the three purely spacial planes of the lattice.
	Expressing this as $h \times 10^{-3}$, we have $h = 11(2),\:8(2)\:9(2)\:,
	5(1)$, for $N_{t}=4,6,8,12$.

	Monopole Wilson loops were calculated using Eq.(\ref{eqn_dual_flux}). These loops are located
	in the purely spacial planes of the $16^{3}\times N_{t}$ lattice.  One dimension of the loop
	can be regarded as a separation $R$, while the other $S$ is a pseudo-time, and by the same 
	arguments
	as used in the usual case of a symmetric lattice, $W_{mon}(R,S)$ should approach $\exp(-SV_{ps}(R))$, as
	$S$ becomes large, where
	$V_{ps}(R)$ is the pseudo-potential.  The values of $V_{ps}(R)$ were determined by fitting $-\ln(W_{mon})$
	to a straight line in $S$, over the interval $S_{min}=R+2$ to $S_{max}=13$.  Then linear-plus-Coulomb fits
	were performed on $V_{ps}(R)$ over the interval $R=2$ to $R=7$.  The coefficient of the linear term in these
	fits gives our estimate of the spacial string tension due to monopoles. We denote the spacial string tension
	deduced from $W_{mon}$ values obtained from Eq.(\ref{eqn_dual_flux}) as $\sigma_{s}(m_{\mu},\bar{m}_{\mu\nu})$.
	For our old results  where the $\bar{m}_{\mu\nu}$ term was omitted in 
	Eq.(\ref{eqn_dual_flux})
	\cite{js_lat94}, we use the
	symbol $\sigma_{s}(m_{\mu})$. In table I , we compare these two determinations of $\sigma_s$ to the
	full $SU(2)$ results  of \cite{bali} on  $32^{3}\times N_{t}$ lattices.
	As can be seen by a glance at the table, the $\sigma_{s}(m_{\mu},\bar{m}_{\mu\nu})$ values agree well with the
	full $SU(2)$ results, whereas the values obtained by omitting the contribution due to $\bar{m}_{\mu\nu}$
	are clearly too small.  

	The values of the $\bar{m}_{\mu\nu}$ are of course highly correlated with the 
	magnetic current in a given configuration.  However, the $\bar{m}_{\mu\nu}$ cannot be constructed from
	knowledge of the magnetic current alone. A useful equation relating $\bar{m}_{\mu\nu}$ to $F^{m*}_{\mu\nu}$ 
	can be derived by summing Eq.(\ref{eqn_dual}) over the $\mu-\nu$ plane.  It is possible to show that
	$\bar{m}_{\mu\nu}$  can be recovered from $F^{m*}_{\mu\nu}$ using this equation only 
	if $|\sum_{x}m_{\mu\nu}| < \frac{1}{2}A_{\alpha\beta}$,
	where $A_{\alpha\beta}$ is the area in lattice units of the plane dual to the $\mu-\nu$ plane. If this restriction is ignored
	and the resulting estimate of $\bar{m}_{\mu\nu}$ is used in Eq.(\ref{eqn_dual_flux}), values of $\sigma_{s}$ 
	which are intermediate between those of columns 1 and 2 of Table I are obtained.  We have also checked
	that using the $\bar{m}_{\mu\nu}$ term by itself  to calculate $W_{mon}$ does not produce an area
	law for Wilson loops, so the effect of $\bar{m}_{\mu\nu}$ is not a simple additive term in the string tension.

	In summary, the  apparent serious discrepancy between monopole and full $SU(2)$ answers for the spacial
	string tension has been removed.  The price paid is that a rather large contribution from Dirac sheets
	must be included.  Dirac sheets are no longer ``invisible" in a finite volume.  It will clearly be
	of great interest to further explore this effect as a function of spacial volume and temperature, including
	zero temperature.

Part of this work was carried out when two of us ( J. Stack and R. Wensley) were
visitors at the University of Wales, Swansea.
This work was supported in part by the National Science Foundation under
Grant No. NSF PHY 94-12556 and Grant No. NSF PHY 94-03869, 
and the Higher Education Funding Council for Wales~(HEFCW).  
The calculations were carried out on the Cray
C90 system at the San Diego Supercomputer Center (SDSC),
supported in part by the National Science
Foundation.


\begin{thebibliography}{99}
\bibitem{gribov} We take one MAG gauge-fixed configuration for each original
$SU(2)$ configuration. Based on recent work by A. Hart and M. Teper, (to be
published) our conclusion that inclusion of Dirac sheets effects is essential
is likely to survive in a more elaborate calculation involving several 
gauge copies for each configuration.
\end{thebibliography}
\end{document}